# FOURIER-RAMAN SPECTRA OF ALKALI METAL AND THALLIUM HYDROGEN PHTHALATE SINGLE CRYSTALS

B. N. Mavrin[*]

*Institute of Spectroscopy RAS, Troitsk, Moscow region, Russia 142190*

M. V. Koldaeva[**], R. M. Zakalukin, T. N. Turskaya

*Institute of Crystallography RAS, Leninskii prospect 59, Moscow, Russia 119333*

(Dated: April 21, 2005)

Polarized Fourier-Raman spectra of K, Rb and Tl hydrogen phthalates are studied in the 220–3300 cm$^{-1}$ region at the 1.064 μm excitation. The frequencies of internal vibrations in hydrogen phthalates are assigned to vibrations of orthophenylene and carboxyl groups. A substitution of K for Rb and Tl gave rise to a small low-frequency shift of vibrations. It is observed the multiband structure of the stretching OH-vibrations due to the fermi-resonance interactions. A number of additional bands are found in the spectra of deuterium potassium acid phthalate. It is suggested that only the partial substitution of hydrogen atoms for deuterium takes place in both orthophenylene and carboxyl groups.

## 1. INTRODUCTION

Alkali metal hydrogen phthalate (MAP) crystals, M(C$_6$H$_4$COOH · COO), are widely known for their application in the long-wave X-ray spectrometers [1]. Their optical, piezoelectric and elastic properties are investigated in detail [2–4]. Recently, MAP crystals were used as substrates for a deposition of thin films of organic nonlinear materials [5].

MAP crystals are crystallized as noncentrosymmetric or centrosymmetric rhombic structures depending on the cation. In particular, MAP crystal with M = K, Rb, and Tl are noncentrosymmetric crystals. These molecular crystals are characterized by a presence of various bonds: covalent (inside anions [C$_6$H$_4$COOH·COO]$^-$), ionic (cation–anion), Van-der-Waals (between chains of anions), and intermolecular hydrogen bonds O-H⋯O (H-bonds between anions in chains). In these crystals H-bonds are very short (~2.5 Å) and, hence, they may be attributed to strong H-bonds for which are possible the fermi-resonance interactions of stretching vibrations ν(O-H) with sum combinations of bending in-plane (β (O-H)) and out-of-plane (γ(O-H)) vibrations owing to intraanionic anharmonicities [6].

In the present paper we report results of the fourier-Raman study of MAP crystals (M = K, Rb, and Tl) in different scattering geometries. Earlier the

---

[*] E-mail: Mavrin@isan.troitsk.ru
[**] E-mail; mkoldaeva@ns.crys.ras.ru

Raman spectra at 0.5145 μm excitation [7], IR spectra [7, 8] and neutron scattering [9] of potassium hydrogen phthalate (KAP) and deuterium KAP (DKAP) were obtained.

## 2. EXPERIMENTAL

The potassium, rubidium (RbAP) and thallium (TlAP) hydrogen phthalate crystals are water-soluble. They were grown from water solution by a technique of lowering temperature from 46 to 30° C at an intense mixing in darken conditions [2, 3]. DKAP was grown in similar way, but starting KAP was dissolved in $D_2O$.

KAP, RbAP, and TlAP are the isostructural crystals [10–12]. Structural model of KAP crystals and the unit cell [10] is shown in Fig. 1. In the parallel planes (010) the corrugated layers of the $M^+$ cations are situated. The anions consist of the phenylene and carboxyl groups and they are located by double layers between the cation layers. The groups $-COO^-$ form H-bonds with the carboxyl group –COOH of the nearest anion along the $c$ axis [10]. H-bonds are marked by the dash line in Fig. 1.

The structure of KAP is classified as the orthorhombic system with the space group $Pca2_1$ ($C_{2v}^5$, $Z = 4$). The crystallographic axes were found by etching of the cleavage plane (010) with water. The corners of the etched figures pointed to the face of the rapid-growing pyramid that corresponded to the negative coming out of the $c$ axis [10]. The direction [001] is the native crystalline edge, the direction [100] lies in the cleavage plane perpendicularly to the direction [001], and the cleavage plane (010) is the native plane of pinacoid. The oriented crystals were cut with the water saw and polished on the wet silk.

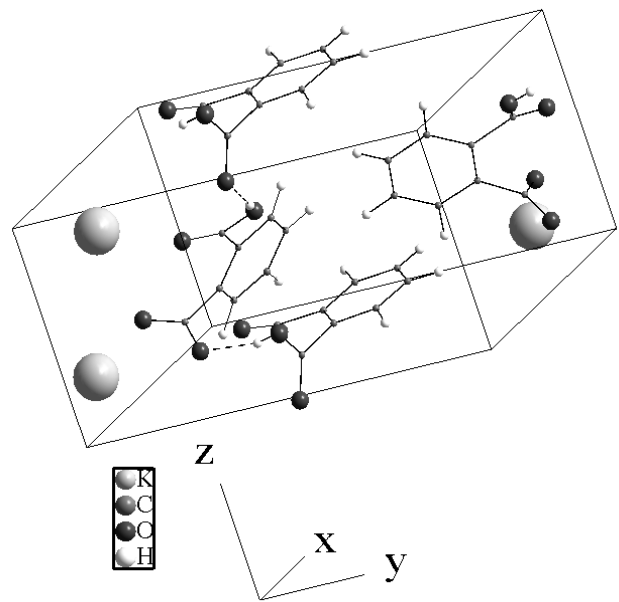

**Fig. 1.** The structural model of the KAP crystal and unit cell.

The X-ray diffraction was studied with the diffractometer Shimadzu XRD-600 (Cu $K_\alpha$ radiation, Ni filter). The diffractograms were obtained with the 0.01° steps and the 1 s exposition. The cell parameters with the $(0.5–1) \times 10^{-3}$ Å accuracy were found after the full-profile analysis of diffractograms with the GSAS program package. The measured and literary parameters of the MAP unit cells are given in Table 1.

The Raman spectra in the $-500 \div 3300$ cm$^{-1}$ region were obtained by the Fourier-Raman spectrometer RFS-100/S [13] at the 1.064 μm laser excitation with power 100 mW and the 4 cm$^{-1}$ resolution at room temperature. The study was carried out in the backscattering geometry with the polarization analysis of the scattered light.

**Table 1.** The unit cell parameters of the MAP crystals (in Å)

| Axis | Crystals | | | | |
|---|---|---|---|---|---|
| | KAP | DKAP | KAP [8] | RbAP[*] [9] | TlAP[*] [10] |
| $a$ | 9.624(6) | 9.630(7) | 9.614(4) | 10.064(2) ($b$) | 10.047(2) ($b$) |
| $b$ | 13.333(1) | 13.340(1) | 13.330(1) | 13.068(2) ($c$) | 12.878(2) ($c$) |
| $c$ | 6.483(0) | 6.485(1) | 6.479(4) | 6.561(1) ($a$) | 6.615(2) ($a$) |

[*] In brackets the axes are given according to references.

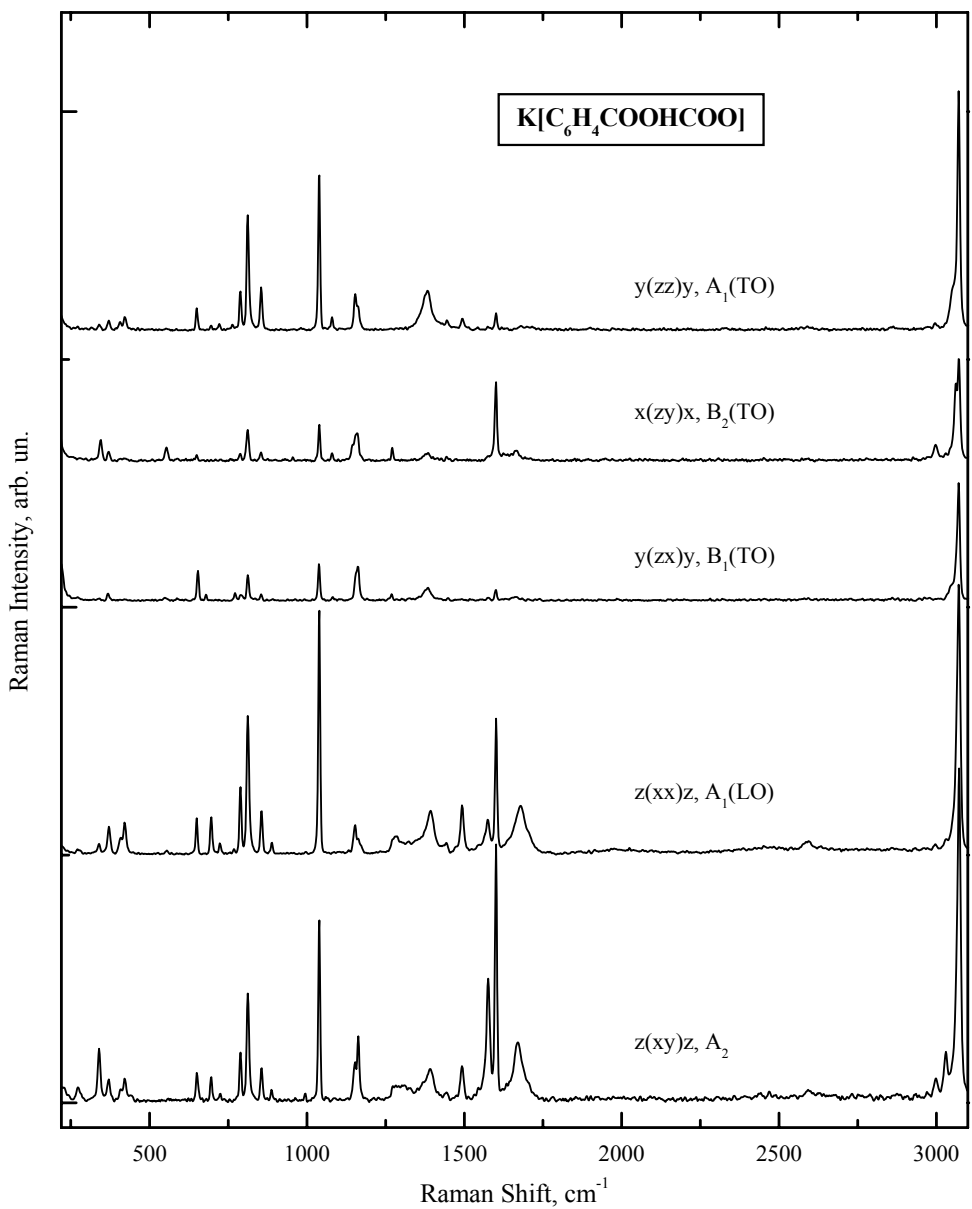

**Fig. 2.** The polarized Raman spectra of the KAP crystal.

## 3. THE DISCUSSION OF DATA

According to X-ray data [10–12], all atoms of the MAP formula unit (hereafter 'molecule') are located in the general position. Therefore all $3N-6$ internal vibrations, where $N$ is the number of atoms in molecule ($N = 18$), should be active in the Raman spectra. Besides, since the MAP unit cell contains four molecules, each vibration is split into four Davydov's components of the symmetry $A_1$, $A_2$, $B_1$ and $B_2$, which are also Raman active.

It is expedient to divide the vibrational representation of MAP into the vibrations of the orthophenylene (OPh) and carboxyl groups. The frequencies and forms of 30 internal vibrations of OPh-group are known [14]: 15 stretching (C–H, C=C, C–C', skeletal), 6 bending in-plane (C–H, C–C'), and 9 bending out-of-plane (C–H, C–C', skeletal) vibrations, where C' is the carbon atom substituting the hydrogen atom in the benzene ring. In view of a weak (ionic) bond of the metal atoms with anions, it is difficult to expect an appearance of vibrations of the metal atoms in the Raman spectra. However, it is possible an influence of these atoms on the spectra of OPh and carboxyl groups.

**Table 2.** The Raman frequencies of the A$_1$ and A$_2$ symmetry in the KAP, RbAP and TlAP crystals

|    | x(zz)x, A$_1$(TO) | | | z(xx)z, A$_1$ (LO) | | | z(xy)z, A$_2$ | | | Assignment |
|----|-----|------|------|------|------|------|------|------|------|------------|
|    | K   | Rb   | Tl   | K    | Rb   | Tl   | K    | Rb   | Tl   |            |
| 1  | 271 | 276  | 268  | 271  |      |      | 273  | 277  | 274  |            |
| 2  | 283 |      | 282  | 280  | 279  | 274  |      |      |      |            |
| 3  | 340 | 342  | 341  | 340  | 340  | 338  | 339  | 340  | 338  |            |
| 4  | 370 | 372  | 369  | 371  | 371  | 369  | 370  | 371  | 368  |            |
| 5  | 406 | 406  | 408  | 409  | 409  | 408  | 408  | 408  | 409  |            |
| 6  | 421 | 420  | 420  | 421  | 420  | 420  | 441  |      | 440  |            |
| 7  | 556 | 557  | 553  | 555  | 558  |      |      | 558  |      |            |
| 8  | 583 | 582  |      |      | 583  |      |      |      |      |            |
| 9  | 649 | 649  | 648  | 650  | 649  | 648  | 651  | 650  | 651  | γ(C=O)     |
| 10 | 695 | 696  | 691  |      |      |      | 696  | 696  | 691  |            |
| 11 | 721 | 721  | 718  | 725  | 722  | 719  | 725  | 722  | 718  |            |
| 12 | 763 | 761  | 760  | 768  | 765  | 762  |      | 767  | 761  | γ(O-D)     |
| 13 | 788 | 790  | 788  | 789  | 792  | 788  | 788  | 791  | 788  | β(C=O)     |
| 14 | 812 | 812  | 809  | 812  | 812  | 807  | 812  | 812  | 807  |            |
| 15 | 855 | 855  | 851  | 856  | 856  | 852  | 856  | 856  | 852  |            |
| 16 |     |      | 897  | 889  | 890  | 890  | 887  | 889  | 888  |            |
| 17 |     |      |      |      |      |      |      |      |      |            |
| 18 |     |      |      |      |      |      | 994  | 997  | 997  |            |
| 19 | 1039| 1039 | 1040 | 1039 | 1038 | 1039 | 1039 | 1039 | 1039 |            |
| 20 |     |      |      |      |      |      |      |      |      | β(O-D)     |
| 21 | 1079| 1079 | 1078 |      |      |      |      |      |      |            |
| 22 |     |      | 1087 |      |      |      |      |      |      | γ(O-H)     |
| 23 | 1153|      | 1152 | 1153 | 1152 | 1147 | 1153 | 1151 | 1148 |            |
| 24 | 1162| 1162 | 1161 | 1164 | 1162 | 1163 | 1163 | 1161 | 1162 |            |
| 25 |     |      | 1198 |      |      |      |      |      |      |            |
| 26 |     | 1246 |      |      |      |      |      |      |      |            |
| 27 | 1268| 1269 | 1269 | 1282 | 1272 | 1274 | 1274 | 1269 | 1270 | ν(C-OH)    |
| 28 |     |      | 1302 | 1323 | 1287 | 1285 | 1329 | 1315 | 1301 |            |
| 29 | 1382| 1386 | 1386 | 1393 | 1395 | 1395 | 1392 | 1393 | 1391 |            |
| 30 | 1444| 1445 | 1445 | 1444 | 1443 | 1444 | 1444 | 1443 | 1444 | β(O-H)     |
| 31 |     |      |      |      |      |      |      |      |      |            |
| 32 | 1494| 1493 | 1491 | 1493 | 1493 | 1491 | 1493 | 1492 | 1490 |            |
| 33 | 1517|      |      |      |      |      |      |      |      |            |
| 34 | 1542| 1545 | 1552 | 1545 |      |      |      | 1548 | 1541 |            |
| 35 | 1574|      | 1574 | 1575 | 1574 | 1577 | 1576 | 1575 | 1575 |            |
| 36 | 1601| 1599 | 1598 | 1601 | 1599 | 1600 | 1601 | 1599 | 1598 |            |
| 37 | 1679| 1622 | 1670 | 1679 | 1676 | 1677 | 1670 | 1671 | 1664 | ν(C=O)     |
| 38 |     |      |      |      |      |      |      |      |      |            |

The polarized Raman spectra of KAP are shown in Fig. 2, and the Raman frequentcies of KAP, RbAP, TlAP, and DKAP are collected in Tables 2 and 3. The frequencies below 220 cm$^{-1}$ are likely corresponded to external vibrations of molecules and they will be discussed elsewhere. We shall use a set of the characteristic frequencies for di-orthosubstituted benzene [14] to assign the Raman bands.

**Table 3.** The Raman frequencies of the $B_1$ and $B_2$ symmetry in the KAP, RbAP and TlAP crystals and the $A_1$, $B_1$ and $B_2$ symmetry in the DKAP

|  | y(zx)y, $B_1$(TO) | | | x(zy)x, $B_2$(TO) | | | DKAP | | | Assignment |
|---|---|---|---|---|---|---|---|---|---|---|
|  | K | Rb | Tl | K | Rb | Tl | $A_1$(TO) | $B_1$(TO) | $B_2$(TO) |  |
| 1 | 271 |  |  |  |  |  | 267 | 267 | 268 |  |
| 2 |  | 285 | 279 |  |  | 279 |  |  |  |  |
| 3 | 340 | 343 | 341 | 344 | 344 | 339 | 338 | 338 | 340 |  |
| 4 | 368 | 364 | 363 | 370 | 368 | 368 | 368 | 367 | 368 |  |
| 5 |  | 406 |  | 407 | 406 | 404 | 401 | 401 | 409 |  |
| 6 | 420 | 424 |  | 421 | 418 | 420 | 419 | 420 | 420 |  |
| 7 | 549 | 551 | 551 | 554 | 556 | 554 | 552 | 548 | 553 |  |
| 8 | 588 | 586 | 583 |  | 588 | 582 |  | 586 |  |  |
| 9 | 653 | 654 | 652 | 650 | 650 | 651 | 641 | 643 | 641,652 | γ(C=O) |
| 10 | 680 | 679 | 677 | 697 | 697 | 692 | 678 | 679 | 679 |  |
| 11 | 720 | 722 |  | 720 | 722 | 717 | 721 | 721 | 721 |  |
| 12 | 771 | 770 | 766 | 752 | 769 | 766 | 754 | 754 | 761 | γ(O-D) |
| 13 | 791 | 792 | 789 | 788 | 791 | 788 | 775 | 778 | 775 | β(C=O) |
| 14 | 812 | 812 | 809 | 812 | 812 | 808 | 807 | 808 | 809 |  |
| 15 | 855 | 855 | 852 | 854 | 855 | 852 | 854 | 853 | 852 |  |
| 16 | 892 | 895 | 895 |  |  |  | 888 | 891 | 888 |  |
| 17 | 963 | 965 | 965 | 955 | 958 | 949 | 961 | 961 | 955 |  |
| 18 | 993 | 995 | 994 | 992 | 992 | 999 | 994 | 994 | 994 |  |
| 19 | 1038 | 1039 | 1039 | 1039 | 1040 | 1039 | 1038 | 1038 | 1038 |  |
| 20 |  |  |  |  |  |  | 1058 |  | 1058 | β(O-D) |
| 21 | 1082 | 1082 | 1083 | 1080 | 1080 | 1078 | 1079 | 1082 | 1080 |  |
| 22 |  |  |  |  |  | 1087 |  |  |  | γ(O-H) |
| 23 | 1155 | 1154 | 1154 | 1155 | 1155 | 1152 | 1151 | 1157 | 1154 |  |
| 24 | 1162 | 1162 | 1162 | 1160 |  | 1160 | 1169 | 1163 | 1163 |  |
| 25 | 1191 |  |  | 1200 | 1194 | 1196 |  |  |  |  |
| 26 |  | 1237 |  | 1246 | 1223 | 1235 |  |  |  |  |
| 27 | 1269 | 1268 | 1267 | 1271 | 1270 | 1269 | 1270 | 1269 | 1270 | ν(C-OH) |
| 28 |  | 1293 |  | 1289 | 1293 | 1292 | 1294 | 1289 | 1291 |  |
| 29 | 1382 | 1385 | 1385 | 1382 | 1384 | 1386 | 1378 | 1379 | 1378 |  |
| 30 | 1448 | 1444 | 1444 | 1444 |  | 1445 | 1443 | 1444 | 1440 | β(O-H) |
| 31 |  |  | 1474 |  | 1475 | 1470 | 1471 |  |  |  |
| 32 | 1491 | 1492 | 1492 |  | 1495 | 1491 | 1489 | 1490 | 1490 |  |
| 33 | 1517 |  |  |  |  | 1518 |  |  |  |  |
| 34 | 1554 |  |  | 1545 | 1550 | 1546 |  | 1541 | 1548 |  |
| 35 | 1576 | 1574 | 1576 | 1577 |  | 1576 | 1574 | 1574 | 1575 |  |
| 36 | 1600 | 1600 | 1598 | 1600 | 1599 | 1599 | 1600 | 1600 | 1601 |  |
| 37 | 1663 | 1663 | 1659 | 1669 | 1663 | 1670 |  | 1669 | 1664 | ν(C=O) |
| 38 |  |  |  |  |  |  | 1688 | 1688 | 1692 | ν(C=O) |

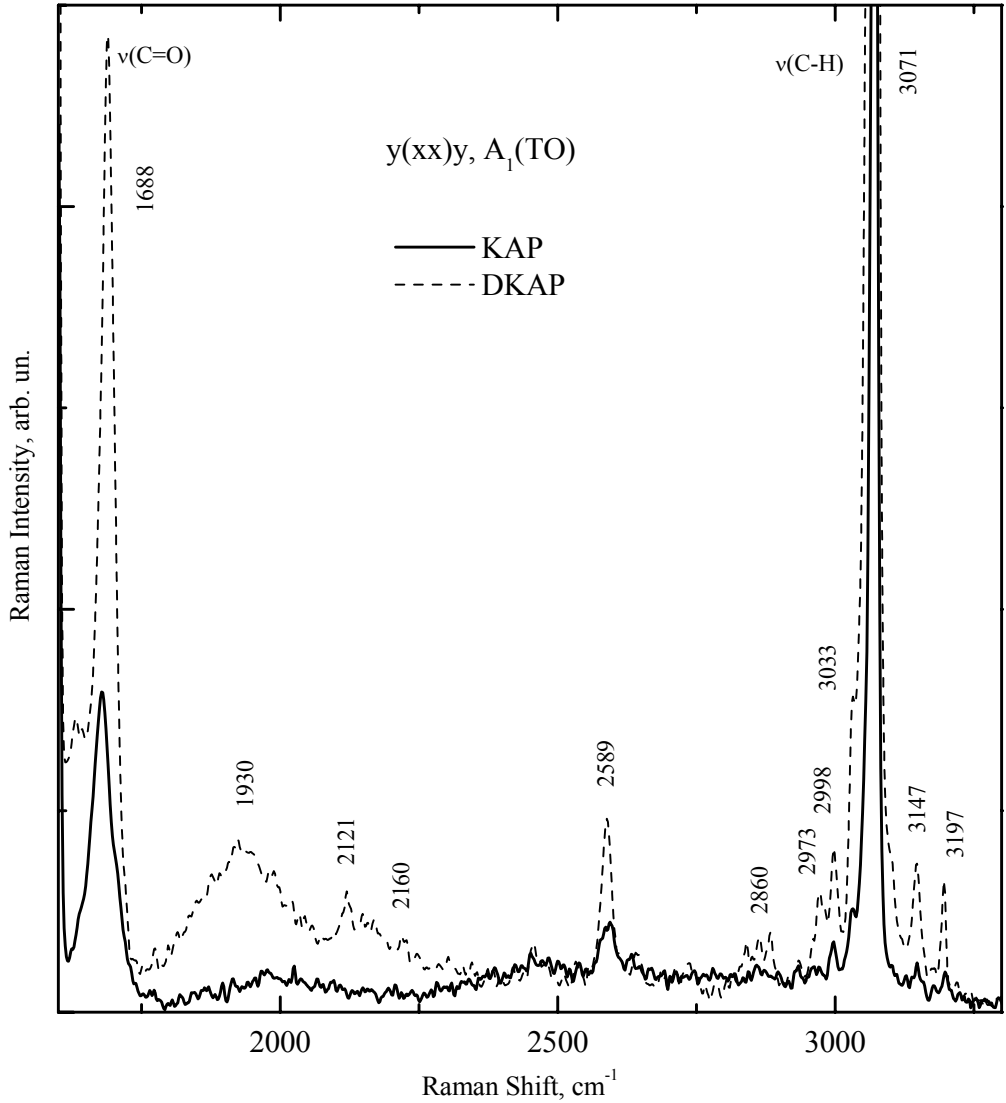

**Fig. 3.** The Raman spectra of the KAP and DKAP crystals in the region of the O-H(D) and C-H(D) vibrations.

It is known, that in case of the strong H-bands the stretching vibrations ν(O-H) are shift in the region below 3000 cm$^{-1}$, as, for example, in the spectra of the KDP crystals [15]. In the Raman spectra of KAP it is seen a weak diffusive scattering in the 1800–3000 cm$^{-1}$ region with the broad bands near 2000 and 2450 cm$^{-1}$, and the less significant shoulder near 2700 cm$^{-1}$, as well as the narrow band near 2590 cm$^{-1}$ (Figs. 2, 3). Earlier [7] the 2590 cm$^{-1}$ band was assigned to the sum combinations of the skeletal vibrations, and the broad bands to vibrations of H-bonds. The emergence of the multiband structure for ν(O-H) may be explainned by the fermi-resonance interactions [6]. The stretching vibrations ν(O-H) were most intense in the y(xx)y scattering geometry when polarizations of the exciting and scattered radiation were aligned along the preferential direction of H-bonds (Fig. 3).

Earlier [9] the bending vibrations β(O-H) and γ(O-H) in KAP were found (near 177 and 138 meV, respectively). We have assigned the Raman bands at 1445 and 1087 cm$^{-1}$ to them that it was in agreement with a suggestion in [7]. Note, that a minimum in diffusive scattering of ν(O-H) near 2200 cm$^{-1}$ (Fig. 3) is close to the 2γ(O-H) overtone that does not contradict to the assumption of the fermi-resonance of this overtone with the ν(O-H) diffusive band. Earlier [6] it is argued that the most strong fermi-resonance interaction takes place with the overtones of the bending out-of-plane bands γ(O-H).

The narrow Raman bands in the 2800–3200 cm$^{-1}$ region (Figs. 2, 3) should be assigned to the ν(C-H) stretching vibrations of the OPh-group [14].

The ν(C=O) stretching vibrations are specified unambiguously. They form the broad band near 1670 cm$^{-1}$. This band split into two components in the A$_1$(TO) spectra of KAP, RbAP and TlAP that may be due to the presence of the different bondlengths C=O in the MAP crystals. The frequencies of the bending vibrations β(C=C) and γ(C=C) may be identified less precisely, although their possible positions for the COOX substitutions in benzene are known [14]: 700–825 and 525–695 cm$^{-1}$, respectively.

In the spectra of orthophtalic acid with two group-substituents COOH in benzene the 1282 cm$^{-1}$ band is assigned to the stretching vibration ν(C-OH). Its analogue in the MAP spectra is also present in this region and it is most intense in the z(xx)z due to the preferential direction of the C-OH bond along the *a* axis.

The other bands of the intraanionic vibrations in the region below 1600 cm$^{-1}$ may be assigned to the internal vibrations of OPh-group. Their frequencies weakly depend on substituents and to assign bands one can follow to the known data for OPh-groups [14]. Then, five high-frequency bands (1494, 1517, 1542, 1574, and 1601 cm$^{-1}$ in the KAP spectra) may be assigned to the stretching vibrations ν(C-C). However, the assignment of other bands is less reliable because of their proximity and a possible small deviation from proposed order in [14].

If deuterium enters in the carboxyl group of MAP, then one can expect a change of the distance between oxygen atoms forming H-bond and, as consequence, a change of the cell parameters. We have found that the cell parameters in DKAP increased by 0.6, 0.5, and 0.3 % for the *a*, *b* and *c* axis, respectively, in comparison with KAP. These changes may evidence that deuterium substitutes hydrogen in the carboxyl groups.

The frequencies of DKAP were slightly changed in comparison with those in KAP. All frequency shifts were with lowering frequencies in DKAP. Note, that the band ν(C=O) in DKAP was split not only in the A$_1$(TO) spectra, but also in the B$_1$ and B$_2$ spectra.

In the DKAP spectra new bands were appeared in the region of the stretching vibrations ν(O-H) and ν(C-H). Assuming that the bands near 3000-3070 cm$^{-1}$ are related to the ν(C-H) vibrations, one can anticipate the ν(C-D) in the 2120–2170 cm$^{-1}$ region. Really, the narrow band at 2121 cm$^{-1}$ is seen in the DKAP spectra (Fig. 3) that evidence a substitution of H for D in the OPh-group.

It is of interest the emergence of the broad band at 1930 cm$^{-1}$ in the DKAP spectra (Fig. 3). It may be assigned to the ν(O-D) stretching vibrations that is an analogue of the 2000 cm$^{-1}$ band in KAP. Earlier [15] in this region the ν(O-H) and ν(O-D) bands were observed in spectra of the KDP and DKDP crystals. The presence of the stretching vibrations ν(O-H(D)) and ν(C-H(D)) in the DKAP spectra is the evidence of the partial deuteration of KAP at the given technique of the DKAP growth.

## 4. CONCLUSIONS

We have studied the polarized Raman spectra of the potassium, rubidium and thallium hydrogen phthalates as well as the potassium deuterium phthalate. The spectra in the different scattering geometries are allowed to obtain the vibrations of the A$_1$, A$_2$, B$_1$, and B$_2$ symmetry. It is shown that the observed bands can be assigned to vibrations of the different structural groups: the orthophenylene group and the bonds C=O, C-H, C-OH and O-H. The ν(O-H) stretching vibrations forms the diffusive band from 1800 to 3000 cm$^{-1}$ that split into a few components due to the fermi-resonance interactions with the overtone and sum combinations of bending vibrations of H-bonds. The change potassium for rubidium or thallium gives rise to a small low-frequency shift of bands. We have found a number of the additional bands in the spectra of potassium deuterium phthalate that give evidence a partial substitution of hydrogen for deuterium occurs in both the orthophenylene and carboxyl groups.

This study was supported partially by the Russian Foundation for Basic research, grant no. 03-02-17021. Authors are also grateful to B. A. Kolesov, V. V. Kravchenko, and A. V. Okotrub for for the help in the beginning of work.